\documentclass[prl,nofootinbib,superscriptaddress,twocolumn]{revtex4}

\usepackage{amsfonts,amssymb,amsthm,bbm}

\usepackage{amsmath}

\usepackage{hyperref}

\usepackage{color,psfrag}
\usepackage[dvips]{graphicx}

\usepackage{tikz}
\usetikzlibrary{calc}
\usetikzlibrary{decorations.pathmorphing}
\usetikzlibrary{shapes.geometric}
\usetikzlibrary{arrows,decorations.markings}

\newcommand{\C}{{\mathbb C}}
\newcommand{\N}{{\mathbb N}}
\newcommand{\R}{{\mathbb R}}

\newcommand{\cG}{{\mathcal G}}

\newcommand{\cL}{{\mathcal L}}

\newcommand{\cV}{{\mathcal V}}

\newcommand{\cI}{{\mathcal I}}
\newcommand{\cZ}{{\mathcal Z}}

\newcommand{\SU}{\mathrm{SU}}

\newcommand{\be}{\begin{equation}}
\newcommand{\ee}{\end{equation}}
\newcommand{\beq}{\begin{eqnarray}}
\newcommand{\eeq}{\end{eqnarray}}
\newcommand{\bes}{\begin{eqnarray}}
\newcommand{\ees}{\end{eqnarray}}

\newcommand{\blue}{\color{blue}}

\newcommand{\alink}[4]
{\draw[decoration={markings,mark=at position 0.6 with {\arrow[scale=1.5,>=stealth]{>}}},postaction={decorate}] (#1) -- node[#3,pos=.5]{$#4$}(#2)}
\newcommand{\link}[2]
{\draw[decoration={markings,mark=at position 0.6 with {\arrow[scale=1.5,>=stealth]{>}}},postaction={decorate}] (#1) --(#2)}
\newcommand{\blink}[2]
{\draw[red, very thick,decoration={markings,mark=at position 0.6 with {\arrow[scale=1.5,>=stealth]{>}}},postaction={decorate}] (#1) --(#2)}
\newcommand{\bllink}[2]
{\draw[blue, decoration={markings,mark=at position 0.6 with {\arrow[scale=1.5,>=stealth]{>}}},postaction={decorate}] (#1) --(#2)}

\def\centerarc[#1](#2)(#3:#4:#5)
{ \draw[#1] ($(#2)+({#5*cos(#3)},{#5*sin(#3)})$) arc (#3:#4:#5); }

\def\centerarcnodes[#1](#2)(#3:#4:#5)(#6,#7)
{\coordinate(#6) at ($(#2)+({#5*cos(#3)},{#5*sin(#3)})$);
	\coordinate(#7) at ($(#2)+({#5*cos(#4)},{#5*sin(#4)})$);
	\draw[#1] ($(#2)+({#5*cos(#3)},{#5*sin(#3)})$) arc (#3:#4:#5); }

\def\angcircle(#1)(#2)(#3:#4) {\coordinate(#1) at ($(#2)+({#4*cos(#3)},{#4*sin(#3)})$); }

\newcommand{\ra}{\rangle}

\newcommand{\f}{\frac}

\def\nn{\nonumber}
\def\pp{\partial}

\def\rd{\mathrm{d}}

\def\vphi{\varphi}
\def\eps{\epsilon}
\def\om{\omega}

\def\tphi{\tilde{\phi}}
\def\tJ{\tilde{J}}
\def\tj{\tilde{j}}
\def\bGamma{\overline{\Gamma}}

\begin{document}

\title{2d Ising Critical Couplings from Quantum Gravity}

\author{{\bf Etera R. Livine}}\email{etera.livine@ens-lyon.fr}
\affiliation{Universit\'e de Lyon, ENS de Lyon, CNRS, Laboratoire de Physique LPENSL, 69007 Lyon, France}
\author{{\bf Valentin Bonzom}}\email{valentin.bonzom@univ-eiffel.fr}
\affiliation{LIGM, CNRS UMR 8049, Université Gustave Eiffel, Champs-sur-Marne, France}

\date{\today}

\begin{abstract}

Using an exact holographic duality formula between the inhomogeneous 2d Ising model and 3d quantum gravity, we provide a formula for "real" zeroes of the 2d Ising partition function on finite trivalent graphs in terms of the geometry of a 2d triangulation embedded in the three-dimensional Euclidean space. The complex phase of those zeroes is given by the dihedral angles of the triangulation, which reflect its extrinsic curvature within the ambient 3d space, while the modulus is given by the angles within the 2d triangles, thus encoding the intrinsic geometry of the triangulation. Our formula can not cover the whole set of Ising zeroes, but we conjecture that a suitable complexification of these "real" zeroes  would provide a more thorough formula.  Nevertheless, in the thermodynamic limit, in the case of flat planar 2d triangulations, our Ising zeros' formula gives the critical couplings for isoradial graphs, confirming its generality.
Finally, the formula naturally extends to graphs with arbitrary valence in terms of  geometry of circle patterns embedded in 3d space.
This approach shows an intricate, but precise, new relation between statistical mechanics and quantum geometry.

\end{abstract}

\maketitle


%
%
%
%
%
%


In the realm of quantum gravity,
as geometry becomes quantum at the fundamental scales, and as classical geometry is meant to emerge at larger scales from quantum correlations and information flow, there is now an increasing interface between statistical physics, condensed matter and research in quantum gravity. Growing on the physics of entanglement and the essential role of quantum information at the heart of the dynamics of quantum geometry, this interface, primarily concerned with the investigation of the phase diagram of the quantum space-time and black hole entropy, resonates today with the theme of holographic dualities.
In this context, we would like to open a new chapter of this interface based on an exact holographic duality recently discovered between 3d quantum gravity and the inhomogeneous 2d Ising model, and show that quantum gravity provides the Ising critical couplings with a natural geometric interpretation and an explicit new parametrization in terms of 2d triangulated surfaces embedded in the 3d space.

We will work with 3d quantum gravity formalized as the Ponzano-Regge state-sum \cite{PR1968,Freidel:2004vi,Barrett:2008wh,Livine:2021sbf}. This discrete path integral defines a topological quantum field theory (TQFT) and provides us with transition amplitudes between quantum states of geometry.
It is indeed the archetype for spinfoam models for quantum gravity path integrals, e.g. \cite{Livine:2024hhc,Rovelli:1993kc,Perez:2012wv}.
Furthermore, it also shown to equivalent to the quantization of 3d gravity as a Chern-Simons theory \cite{Witten:1988hc,Freidel:2004nb} and is a special case of Turaev-Viro invariants \cite{Turaev:1992hq,Turaev:2010pp}. As a topological invariant, all the local information is a priori projected onto the boundary making it a naturally-holographic theory. 

This idea was turned recently into a concrete duality for finite distance holography \cite{Bonzom:2015ova}, where the 3d quantum gravity amplitudes for a bounded region with suitable boundary coherent states are shown to be equal to the inverse square of the inhomogeneous 2d Ising partition function on the boundary. This can be formulated as the supersymmetric localization of a path integral, between some fermionic boundary degrees of freedom (the Ising spins) and some bosonic degrees of freedom, namely the fluctuations of the boundary geometry. Remarkably, this duality comes from a {different} path than the AdS/CFT correspondence and more standard asymptotic holographic formulas, as developed in e.g. \cite{Castro:2011zq,Jian:2019ubz}. These usually rely on a conformal field theory construction on the boundary in a critical regime, while our approach holds exactly, even away from criticality.

Due to this inverse relation between the Ponzano-Regge amplitudes and the 2d Ising partition function, the \emph{zeroes} of the latter, known as Fisher zeroes, become the \emph{poles} of the quantum gravity path integral, and are thus mapped onto asymptotical problems of convergence of series, which can be analyzed by saddle point techniques.
Using results on spin network asymptotics from the well-studied semi-classical limit of the Ponzano-Regge state-sum, we find a new formula for the zeroes of the 2d Ising partition function in terms of \emph{embedded} 2d triangulations. These zeroes are given by complex couplings, whose moduli are given in terms of the \emph{intrinsic} geometry of the triangulations (the 2d angles within each triangle), while their phases are given in terms of the \emph{extrinsic} geometry (the 3d dihedral angles between the planes of neighboring triangles).
In the thermodynamical limit, these zeroes become critical couplings and our formula is compatible with previous works from the mathematical literature on the 2d Ising model on isoradial graphs, e.g. \cite{Boutillier_2018}.

This short work should be considered as part of a larger program interplaying between quantum gravity and statistical physics methods: on the one hand,
investigate the use of geometric saddle point techniques for quantum gravity to provide boundary statistical physics models with a geometrical interpretation, as we do here, and on the other hand, use statistical physics methods to work out the coarse-graining and renormalization flow of the quantum geometry dynamics and explore the phase diagram and the continuum limit of quantum gravity models.

The paper is organized as follows. We first review the duality formula  \cite{Bonzom:2015ova}. We recall the basics of the 2d Ising model and of boundary spin network states of quantum gravity.  We introduce a specific class of coherent states, such that the quantum gravity amplitudes are the inverse square of the 2d Ising partition function. Then a second section is dedicated to deriving a formula for zeros of the Ising partition function from the semi-classical regime of the Ponzano-Regge state-sum and providing them with a parametrization in terms of 2d triangulations.
We derive, what we refer to, as "real" zeroes and conjecture that complexifying our formula should provide a more general ansatz.
We further extend our analysis from 3-valent graphs, dual to 2d triangulations, to graphs with nodes of arbitrary higher valence, in which case our geometric formula for Ising zeros is written in terms of circle patterns embedded in 3d space.
We conclude with a comparaison with known results on Ising zeroes and critical couplings, and discuss possible extensions of the present work.

\section{Quantum Gravity - Ising duality}

The Westbury formula, originally derived in \cite{Westbury1998}, identifies the partition function of the inhomogeneous 2d Ising model as a generating functional for spin network evaluations. Re-derived by Bonzom, Costantino and Livine (BCL) in \cite{Bonzom:2015ova}, this formula was re-interpreted as the equality between the 3d quantum gravity amplitude on the 3-ball with a quantum boundary state (on its 2-sphere boundary) and the squared-inverse of the partition function of an inhomogeneous 2d Ising model living on that boundary. This BCL duality is the example of an exact holographic duality at finite distance, between 3d quantum gravity and a non-critical conformal field theory. It is our starting point for deriving a new formula for the zeroes of 2d Ising model in terms of 2d triangulations.

Let us start with recalling the definition of the 2d Ising model.
Consider a closed planar oriented (connected) graph $\Gamma$, with $N$ nodes and $L$ links.
The  partition function of the Ising model is given by 
\bes
\cI_\Gamma(\{y_\ell\})
&=&
\sum_{\{\sigma_{n}=\pm\}}\prod_{\ell}e^{y_{\ell}\sigma_{s(\ell)}\sigma_{t(\ell)}}
\,, 
\ees
where $n$ and $\ell$ respectively label the nodes and links of the graph, and $s(\ell)$ and $t(\ell)$ denote the source and target nodes of the link $\ell$, as depicted on fig.\ref{fig:graph}.
The link coupling $y_\ell$ is related to the Ising coupling $J_\ell$ by the inverse temperature $\beta$ as follows, $y_\ell=\beta J_\ell$, so that the coefficients  $e^{y_\ell}$ stand for the Boltzmann weights for each link $\ell$.
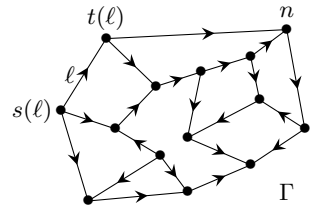
\begin{figure}[h!]

\centering
\begin{tikzpicture}[scale=1.2]

\coordinate(a) at (0,0.2) ;
\coordinate(b) at (.5,1);
\coordinate(c1) at (.6,0);
\coordinate(c2) at (1.1,-0.3);
\coordinate(d) at (.3,-.8);
\coordinate(e1) at (1.05,.47);
\coordinate(e2) at (1.55,.63);
\coordinate(f1) at (2.5,1.1);
\coordinate(f2) at (2.1,.8);
\coordinate(g2) at (1.4,-0.1);
\coordinate(g1) at (2.2,.32);
\coordinate(h) at (2.7,0);
\coordinate(i) at (2.1,-.4);
\coordinate(j) at (1.4,.-.7);

\draw (a) node {$\bullet$} node[left]{$s(\ell)$};
\draw (b) node {$\bullet$} node[above]{$t(\ell)$};
\draw (c1) node {$\bullet$} ;
\draw (c2) node {$\bullet$} ;
\draw (d) node {$\bullet$};
\draw (e1) node {$\bullet$};
\draw (e2) node {$\bullet$};
\draw (f1) node {$\bullet$}++(-0,0.2)node{$n$};
\draw (f2) node {$\bullet$};
\draw (g2) node {$\bullet$};
\draw (g1) node {$\bullet$};
\draw (h) node {$\bullet$} ++(-0.2,-0.7) node{$\Gamma$};
\draw (i) node {$\bullet$};
\draw (j) node {$\bullet$};

\alink{a}{b}{left}{\ell};
\link{a}{c1};
\link{a}{d};
\link{c2}{d};
\link{b}{e1};
\link{b}{f1};
\link{e1}{e2};
\link{e2}{f2};
\link{c1}{e1};
\link{c2}{c1};
\link{e2}{g2};
\link{f2}{g1};
\link{f1}{h};
\link{f2}{f1};
\link{g1}{g2};
\link{h}{g1};
\link{h}{i};
\link{g2}{i};
\link{j}{i};
\link{d}{j};
\link{c2}{j};
						
				
\end{tikzpicture}

\caption{
Planar oriented 3-valent graph $\Gamma$.
}
\label{fig:graph}
\end{figure}
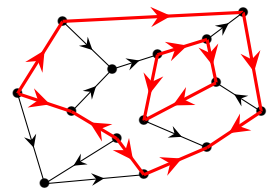
\begin{figure}[h!]

\centering
\begin{tikzpicture}[scale=1.2]

\coordinate(a) at (0,0.2) ;
\coordinate(b) at (.5,1);
\coordinate(c1) at (.6,0);
\coordinate(c2) at (1.1,-0.3);
\coordinate(d) at (.3,-.8);
\coordinate(e1) at (1.05,.47);
\coordinate(e2) at (1.55,.63);
\coordinate(f1) at (2.5,1.1);
\coordinate(f2) at (2.1,.8);
\coordinate(g2) at (1.4,-0.1);
\coordinate(g1) at (2.2,.32);
\coordinate(h) at (2.7,0);
\coordinate(i) at (2.1,-.4);
\coordinate(j) at (1.4,.-.7);

\draw (a) node {$\bullet$} ;
\draw (b) node {$\bullet$} ;
\draw (c1) node {$\bullet$} ;
\draw (c2) node {$\bullet$} ;
\draw (d) node {$\bullet$};
\draw (e1) node {$\bullet$};
\draw (e2) node {$\bullet$};
\draw (f1) node {$\bullet$};
\draw (f2) node {$\bullet$};
\draw (g2) node {$\bullet$};
\draw (g1) node {$\bullet$};
\draw (h) node {$\bullet$} ;
\draw (i) node {$\bullet$};
\draw (j) node {$\bullet$};

\blink{a}{b};
\blink{a}{c1};
\link{a}{d};
\link{c2}{d};
\link{b}{e1};
\blink{b}{f1};
\link{e1}{e2};
\blink{e2}{f2};
\link{c1}{e1};
\blink{c2}{c1};
\blink{e2}{g2};
\blink{f2}{g1};
\blink{f1}{h};
\link{f2}{f1};
\blink{g1}{g2};
\link{h}{g1};
\blink{h}{i};
\link{g2}{i};
\blink{j}{i};
\link{d}{j};
\blink{c2}{j};
				
\end{tikzpicture}

\caption{
Example of even subgraph in {\bf {\color{red} red bold}} of the 3-valent graph $\Gamma$, as a disjoint union of cycles of links.
}
\label{fig:evensubgraph}
\end{figure}

\noindent
The high temperature expansion of the Ising partition function is given by:
\bes
\cI_\Gamma(\{y_\ell\})
&=&
\sum_{\{\sigma_{v}=\pm\}}\prod_{\ell}(\cosh y_{\ell}+\sinh y_{\ell}\sigma_{s(\ell)}\sigma_{t(\ell)}) \crcr
&=&
2^{N}\,\left(\prod_{\ell}\cosh y_{\ell}\right)\,
\sum_{\cG\subset \Gamma} \prod_{\ell\in\cG} \tanh y_{\ell}\,, 
\label{Zising}
\ees
where the sum is performed over all even subgraphs $\cG$ of $\Gamma$, i.e such that each node of the subgraph has even valence.
In the following, we will only consider 3-valent graphs $\Gamma$, in which case even subgraphs are simply collections $\cL$ of disjoint cycles, which we also call loops, as illustrated on fig.~\ref{fig:evensubgraph}. We will then refer to the sum above as the loop expansion of the Ising partition function:
\be
\cI_\Gamma(\{y_\ell\})
=
2^{N}\,\left(\prod_{\ell}\cosh y_{\ell}\right)
P_\Gamma\big{[}\{\tanh y_{\ell}\}\big{]},
\ee
where $P_{\Gamma}$ is a polynomial of $L$ variables, attached to the graph's links:
\be
P_\Gamma\big{[}\{x_{\ell}\}\big{]}=\sum_{\cL\subset\Gamma} \,\,\,\prod_{\ell\in \cL} x_\ell\,.
\ee
The zeroes of the Ising partition function are thus given by the zeroes of this loop polynomial.
As we explain below, it turns out that these correspond to the poles of the 3d quantum gravity path integral.

Indeed, the path integral for 3d quantum gravity, for vanishing cosmological constant and Riemannian space-time signature, can be formulated as a formal topological invariant known as the Ponzano-Regge state-sum. It is defined on triangulated 3-manifolds and constructed in terms of Wigner's 6j-symbols from the recoupling theory of $\SU(2)$. It has been extensively studied, in particular its gauge symmetries and their gauge-fixings, which reflect the fact that it does not depend on the details of the triangulation but only on its topology and its boundary data~\cite{Freidel:2004vi,Freidel:2005bb,Barrett:2008wh,Bonzom:2010ar,Bonzom:2010zh,Bonzom:2011br,Barrett:2011qe,Livine:2021sbf}. 

\begin{figure}[ht!]
\begin{center}
\includegraphics[height=30mm]{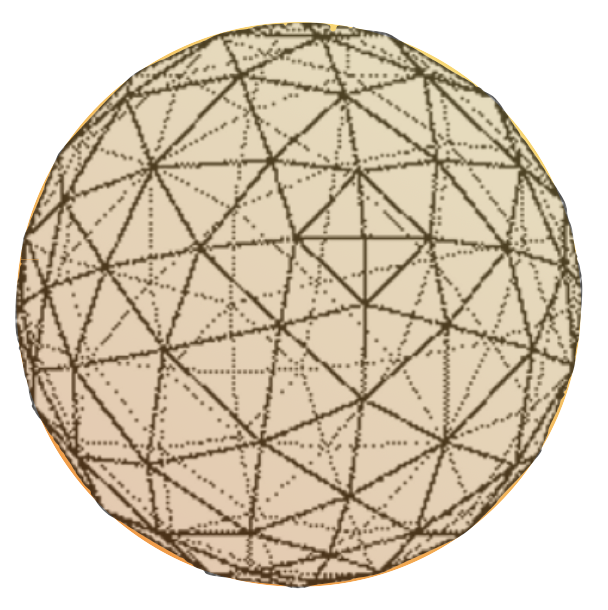}
\caption{
Example of planar triangulation on the 2-sphere. The dual graph is made of nodes dual to the triangles and links dual to the edges of the triangulation.
\label{fig:sphere}}
\end{center}
\end{figure}
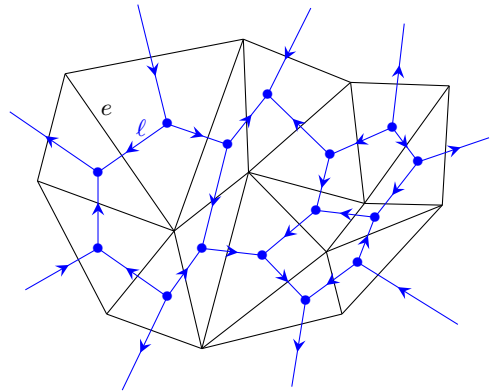
\begin{figure}[ht!]

\centering
	\begin{tikzpicture}[scale=2.3]

		\coordinate (A1) at (-0.39,0.81);
		\coordinate (A2) at (-0.55,0.19);
		\coordinate (A3) at (-0.15,-0.58);
		\coordinate (A4) at (0.4,-0.78);
		\coordinate (A5) at (1.21,-0.58);
		\coordinate (A6) at (2.04,-0.83);
		\coordinate (A7) at (1.79,0.05);
		\coordinate (A8) at (1.83, 0.74);
		\coordinate (A9) at (1.26, 0.76);
		\coordinate (A10) at (0.64, 1.01);
		\coordinate (I1) at (0.24, -0.1);
		\coordinate (I2) at (0.67, 0.24);
		\coordinate (I3) at (1.34, 0.06);
		\coordinate (I4) at (1.12, -0.22);
		
		
		\draw(A1)--(A2);
		\draw(A2)--(A3);
		\draw(A3)--(A4);
		\draw(A4)--(A5);
		\draw(A5)--(A7);
		\draw(A7)--(A8);
		\draw(A8)--(A9);
		\draw(A9)--(A10);
		\draw(A10)--(A1);
		\draw(A1)--(I1);
		\draw(A2)--(I1);
		\draw(A3)--(I1);
		\draw(A4)--(I1); 
		\draw(A4)--(I2); 
		\draw(A4)--(I4); 
		\draw(A7)--(I4); 
		\draw(A8)--(I3);
		\draw(A9)--(I2);
		\draw(I1)--(I2);
		\draw(I2)--(A10);
		\draw(I1)--(A10);
		\draw(I2)--(I3);
		\draw(I3)--(A9);
		\draw(I3)--(A7);
		\draw(I3)--(I4);
		\draw(I4)--(A5);
		\draw(I4)--(I2);    
		
\coordinate (B1) at (0.2, 0.52);
\coordinate (B1b) at (0.55, 0.4);
\coordinate (B2) at (0.78, 0.69);
\coordinate (B2b) at (1.14, 0.34);
\coordinate (B3) at (1.5, 0.5);
\coordinate (B3b) at (1.65, 0.3);
\coordinate (B4) at (1.3, -0.28);
\coordinate (B4b) at (1.4, -0.02);
\coordinate (B5) at (1.06, 0.02);
\coordinate (B6) at (0.75, -0.24);
\coordinate (B6b) at (0.4, -0.2);
\coordinate (B6c) at (1, -0.5);
\coordinate (B7) at (0.2, -0.48); \coordinate (B8) at (-0.2, -0.2); \coordinate (B9) at (-0.2, 0.24); \coordinate (B10) at (0.03, 1.21); \coordinate (B11) at (1.03, 1.19); \coordinate (B12) at (1.57, 1.1); \coordinate (B13) at (2.06, 0.44);
\coordinate (B14) at (1.88, -0.64);
\coordinate (B16) at (0.92, -1.0); \coordinate (B17) at (-0.06, -1.02); \coordinate (B18) at (-0.62, -0.44); \coordinate (B19) at (-0.71, 0.59); 
		
\bllink{B1}{B1b};
\bllink{B1b}{B2};
\bllink{B1}{B9};
\bllink{B10}{B1};
\bllink{B1b}{B6b};
\bllink{B6b}{B6};
\bllink{B6}{B6c};
\bllink{B3}{B2b};
\bllink{B3b}{B13};
\bllink{B3}{B12};
\bllink{B2b}{B2};
\bllink{B2b}{B5};
\bllink{B11}{B2};
\bllink{B3}{B3b};
\bllink{B3b}{B4b};
\bllink{B4}{B4b};
\bllink{B4b}{B5};
\bllink{B4}{B6c};
\bllink{B14}{B4};
\bllink{B5}{B6};
\bllink{B7}{B6b};
\bllink{B6c}{B16};
\bllink{B7}{B8};
\bllink{B7}{B17};
\bllink{B8}{B9};
\bllink{B9}{B19};
\bllink{B18}{B8};

\draw (0.05, 0.5) node{{\blue $\ell$}};
\draw (-0.15, .6) node{{$e$}};
		
\draw[blue] (B1) node{$\bullet$};
		\draw[blue] (B2) node{$\bullet$}; \draw[blue] (B2b) node{$\bullet$}; 
		\draw[blue] (B1b) node{$\bullet$}; 
		\draw[blue] (B3) node{$\bullet$};\draw[blue] (B3b) node{$\bullet$};
		 \draw[blue] (B4) node{$\bullet$}; 
		 \draw[blue] (B4b) node{$\bullet$}; 
		\draw[blue] (B5) node{$\bullet$}; \draw[blue] (B6) node{$\bullet$}; 
		\draw[blue] (B6b) node{$\bullet$}; \draw[blue] (B6c) node{$\bullet$}; 
		\draw[blue] (B7) node{$\bullet$};
		 \draw[blue] (B8) node{$\bullet$}; 
		\draw[blue] (B9) node{$\bullet$};
	\end{tikzpicture}

\caption{
The 3-valent graph $\Gamma$ (in {\blue blue}) and its dual triangulation. The graph nodes are dual to triangles. The graph links are dual triangulation edges. An example is given by the link {\blue $\ell$} and its dual edge $e$. The spin $j_{\ell}$ carried by the link gives the length in Planck unit of its dual edge, $L_{e}=j_{\ell(e)}\,l_{Planck}$.
}
\label{fig:spinnet}
\end{figure}

For a 3-ball, the bulk topology is trivial, and the Ponzano-Regge amplitude entirely depends on the boundary data defining the quantum geometry on its boundary 2-sphere. 
Boundary states are given by {\it spin networks}. A spin network is a closed oriented graph drawn on the 2-sphere, and thus planar, coloured with half-integers $j_{\ell}\in\f \N2$ on its links $\ell$.
A half-integer $j$ corresponds to an irreducible representation of the $\SU(2)$ Lie group, and is thus referred to as a spin. The representation spin $j$ has  dimension $d_{j}=2j+1$, and is spanned by the usual magnetic moment basis:
\be
\cV^{j}=\bigoplus_{-j\le m\le j}\C\,|j,m\ra
\,.
\ee
Here, we will consider, for the sake of simplicity of notation, but without actual limitations of scope, to 3-valent graphs. Such planar graphs are dual to triangulations of the 2-sphere (see fig. \ref{fig:sphere}). As illustrated on fig.\ref{fig:spinnet}, each graph node is dual to a triangle, while each graph link  dual to  an edge of the triangulation.

The spin $j_{\ell}$ on the link $\ell$ is in fact the quantum length (in Planck units) of the dual edge:
\be
L_{e}=j_{\ell(e)}\,l_{\text{Planck}}
\,.
\ee
Then the three spins around a node $n$ give the edge lengths of the triangle that is dual to $n$. Denote $\ell_1^n, \ell_2^n, \ell_3^n$ the three links meeting at $n$, and $j_{\ell_1^n}, j_{\ell_2^n}, j_{\ell_3^n}$ their spins. We associate to $n$ the normalized singlet state $|j_{\ell_1^n}, j_{\ell_2^n}, j_{\ell_3^n}\rangle$ in the tensor product of those three $\SU(2)$ representations (in the canonical basis of this tensor product, the components are the 3jm-Wigner symbols, or equivalently the Clebsh-Gordan coefficients). Then the spin network state is
\be
|\{j_\ell\}_\Gamma\rangle = \bigotimes_n |j_{\ell_1^n}, j_{\ell_2^n}, j_{\ell_3^n}\rangle.
\ee

These spin networks are the basis of the Hilbert space of  states of quantum geometry on the 2-sphere. General boundary states are arbitrary superpositions of spin networks. They define quantum boundary conditions, in the sense that there are the quantum equivalent of classical boundary conditions fixing the boundary 2d metric.

As shown in several works \cite{Freidel:2005bb,Barrett:2011qe,Livine:2021sbf}, the quantum gravity amplitude for the 3-ball with a spin network boundary state is given by the evaluation of the corresponding wave-function on flat connections. Mathematically, this is encoded in the \emph{spin network evaluation} for the colored 3-valent graph $\Gamma$, equipped with an edge orientation $o$. It is given as follows by the contraction of the 3jm-Wigner symbols living at the graph nodes $n$:
\beq
\label{eq:tensorsn}
s_{\Gamma}(\{j_\ell\},o)
&\equiv
\sum_{\{m_\ell\}} &\prod_\ell (-1)^{j_\ell-m_\ell}\\
&&\prod_n \begin{pmatrix}  j_{\ell_{1}^{n}} & j_{\ell_{2}^{n}} & j_{\ell_{3}^{n}}\\ \epsilon_{\ell_{1}} ^{n}m_{\ell_{1}^{n}} & \epsilon_{\ell_{2}}^{n} m_{\ell_{2}^{n}}& \epsilon_{\ell_{3}}^{n} m_{\ell_{3}^{n}} \end{pmatrix}\,,
\nn
\eeq
%
with the  sign $\epsilon_{\ell}^{n}=-1$ resp. $1$ recording if the link $\ell$ is oriented inwards ($n=t(\ell)$) resp. outwards ($n=s(\ell)$). Let us choose a Kasteleyn orientation $o$ \cite{Kasteleyn}, which always exists for a planar graph, then the evaluation is independent of the specific choice of edge orientations \cite{Bonzom:2015ova}. 
We can thus safely drop the $o$ index in the following.

Now, the key to the  duality between the 2d Ising model and 3d quantum gravity is to consider a particular class of superpositions of spin networks on the boundary of the 3-ball. Similarly to coherent states, these superpositions are defined by a set of complex parameters $Y_{\ell}$ associated to the links, which controls the probability distribution of spins on the boundary. The quantum gravity amplitudes for such spin superpositions are, as given in \cite{Bonzom:2015ova}:
\be
\cZ_{\Gamma}(\{Y_\ell\})
\,=\,
\sum_{\{j_\ell\}} s_{\Gamma}(\{j_\ell\}) 
\om_{\Gamma}(\{j_\ell\})
\prod_\ell Y_\ell^{2j_\ell}
\,,
\ee
\be
\textrm{with}\quad
\om_{\Gamma}(\{j_\ell\})
=
\prod_n\sqrt{\frac{ (J_n+1)!}{\prod_{\ell\ni n} (J_n-2j_\ell)!}}
\,,
\nn
\ee
where $J_{n}=j_{\ell_1^n} + j_{\ell_2^n} + j_{\ell_3^n}$ is the sum of the spins living on the links attached to $n$.
These amplitudes can, in turn, be interpreted as generating functions for the spin network evaluations. With the factorial factors in the weight $\om_{\Gamma}(\{j_\ell\})$, this distribution looks very similar to a coupled Poisson distribution for the spins. In fact, these precise factors allow for an exact computation of the amplitudes. This feature has been used in a few other works exploring generating functions for spin recoupling symbols and spinfoam amplitudes \cite{Freidel:2012ji,Bonzom:2012bn,Hnybida:2015ioa}.
Here, the quantum gravity amplitudes greatly simplify to:
\be
\cZ_{\Gamma}(\{Y_\ell\})=P_\Gamma^{-2}[\{Y_\ell\}]
\,,
\ee
in terms of the same loop polynomial used in the high-temperature expansion of  the Ising model, as initially proved by Westbury \cite{Westbury1998}.
In fact, if we identify the quantum gravity boundary couplings and the Ising couplings on the boundary graph, $Y_{\ell}=\tanh y_{\ell}$, then we have an exact duality formula relating the two partition functions:
\be\label{dual}
\cZ_{\Gamma}(\{\tanh y_{\ell}\})\, \cI_{\Gamma}(\{y_{\ell}\})^2
\,=\,
\left(
2^{N}\prod_{\ell\in \Gamma} \cosh y_\ell
\right) ^2
\,.
\ee
This duality was shown to result from a supersymmetry mapping the bosonic degrees of freedom of geometry to the fermionic degrees of freedom of the Ising model \cite{Bonzom:2015ova}. This supersymmetry interestingly allows for non-linear generalizations of the formula above, which have not yet been explored in details.

Here, we would like to exploit the realization of the quantum gravity amplitudes as the squared inverse of the loop polynomial, to identify the zeroes of the 2d Ising partition function $ \cI_{\Gamma}(\{y_{\ell}\})$ as the poles of the quantum gravity partition function $\cZ_{\Gamma}(\{Y_\ell\})$.
Since  the quantum gravity poles are understood as semi-classical geometries, this will lead to geometric formula for Ising zeroes.
Then,  in the thermodynamical limit of infinite graphs, where one expects zeroes to become critical couplings, this would produce a geometric formula for Ising's critical couplings. We confirm this in the case of isoradial  graphs.


\section{Ising zeroes as 2d Triangulations}

The advantage of working with the quantum gravity amplitude $\cZ_{\Gamma}= P_{\Gamma}^{-2}$ compared to the Ising partition function $\cI_{\Gamma}\propto P_{\Gamma}$ is that we have traded a \emph{finite sum over Ising spins} $\sigma_{\ell}=\pm1$ for an \emph{infinite series over half-integer spins} $j_{\ell}\in\f\N2$. This allows us to study the property of the loop polynomial $P_{\Gamma}$ from an asymptotic point of view as we consider the large spin limit $j_{\ell}\rightarrow+\infty$. This asymptotic limit naturally corresponds to the semi-classical regime of quantum gravity since it corresponds to working with lengths much larger than the Planck scale. This should thus lead to an interpretation in terms of classical geometries.

Let us look more closely at the mechanisms at hand. At large spins, the spin network evaluation combined with the factorial weight can be approximated at leading order as:
\be
s_{\Gamma}(\{j_\ell\}) 
\om_{\Gamma}(\{j_\ell\})
\underset{j_{\ell}\gg1}\sim
f(\{j_\ell\})\,e^{\phi(\{j_\ell\})}
\,,
\ee
where both functions $f$ and $\phi$ grow algebraically in the spins $j_{\ell}$. Taking into account the couplings $Y_\ell$, we then look for the stationary points of the shifted exponent, $\pp_{j_{\ell}}\tphi=0$ with
\be
\tphi(\{j_\ell\})
=
\phi(\{j_\ell\})
+\sum_{\ell}2j_{\ell}\ln Y_{\ell}
\,,
\ee
i.e. $\pp_{j_{\ell}}\phi=-2\ln Y_{\ell}$. These stationary points $\{j_\ell^{(0)}\}$ have two key properties:
\begin{itemize}

\item they only exist for specific values of the  couplings $Y_{\ell}$, which we will refer to as {\it admissible values}\,;

\item when they exist, they are {\it scale-invariant}, i.e. the rescaled spin configurations $\{\lambda j_\ell^{(0)}\}$ for arbitrary real factor $\lambda$ are also stationary points.

\end{itemize}
Thus for admissible couplings, having stationary lines (instead of isolated stationary points) leads to a divergence of the sum over spins, so that admissible couplings are poles of the quantum gravity amplitude $Z_{\Gamma}$ and, therefore, zeroes of the Ising partition function $\cI_{\Gamma}$.

\begin{figure}[h!]

\centering
\begin{tikzpicture}[scale=1]

\coordinate(a) at (0,0.4) ;
\coordinate(b) at (1,2);
\coordinate(ab) at (0.5,1.2) ;
\coordinate(c1) at (3.5,-0.7);
\coordinate(c2) at (-3,-0.3);
\coordinate(ac1) at (1.75,-0.15);
\coordinate(ac2) at (-1.5,0.05);
\coordinate(bc1) at (2.25,0.65);
\coordinate(bc2) at (-1,.85);
\coordinate(e1) at (1.8,0.45);
\coordinate(e2) at (-0.9,0.6);
\coordinate(f1) at (3,2.2);
\coordinate(f2) at (-2,2.5);
\coordinate(O) at (0.4,-1);

\draw (e1) node {$\bullet$};
\draw (e2) node {$\bullet$};

\draw (ab)++(-0.45,0) node {$\ell$};
\draw (b)++(-0.1,-.35) node {$e$};
\draw (a)++(-.7,-.4) node {$\Delta$};
\draw (a)++(.8,-.5) node {$\widetilde{\Delta}$};
\draw (c1)++(-.25,.65) node {$\tilde{\vphi}_{\ell}$};
\draw (c2)++(.5,.5) node {$\vphi_{\ell}$};
\draw (O)++(0.2,4.5) node {$\theta_{\ell}$};

\draw[thick,dotted] (e1)--(ab);
\draw[thick,dotted] (e1)--(ac1);
\draw[thick,dotted] (e1)--(bc1);
\draw[thick,dotted, decoration={markings,mark=at position 0.98 with {\arrow[scale=1.2,>=stealth]{>}}},postaction={decorate}] (e2)--(ab);
\draw[thick,dotted] (e2)--(ac2);
\draw[thick,dotted] (e2)--(bc2);

\draw[->, thick,>=stealth](e1)--(f1) ;
\draw[->, thick,>=stealth](e2)--(f2) ;
\draw (a)--(b)  --(c1)--(a);
\draw (b)--(c2)--(a);

\centerarc[](c1)(133:162:.6);
\centerarc[](c2)(12:30:.8);

\centerarc[thick,dashed,<->=0.5](O)(55:120:4.2);

\end{tikzpicture}

\caption{
Neighbouring triangles, sharing the edge $e$ dual to the graph link $\ell$. The triangulation is drawn in full lines, while the graph is drawn in dotted lines. The dihedral angle $\theta_{\ell}$ is the angle between the normal vectors to the two triangles, while the 2d angles $\vphi_{\ell}$ and $\tilde{\vphi}_{\ell}$ are the triangle angles at the vertices opposite to the edge.
\label{fig:neighbours}}

\end{figure}
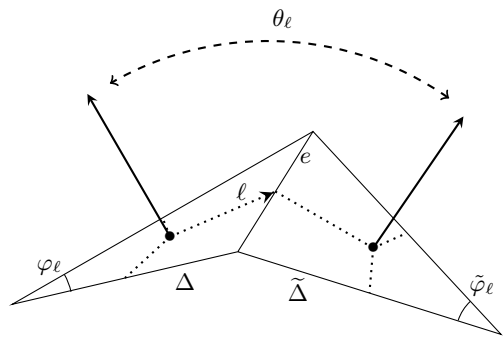

Let us work out the details of those steps.  As we will see below, it leads to the geometric interpretation of the Ising zeroes in terms of the angles of a 2d triangulation embedded in the flat 3d Euclidean space.
Using Stirling formula for the factorials, we get:
\be
\om_{\Gamma}
\underset{j_{\ell}\gg 1}\sim
\prod_{n}
\left(
\f{J_{n}^{J_{n}+\f32}}{2\pi\prod_{\ell\ni n}(J_{n}-2j_{\ell})^{J_{n}-2j_{\ell}+\f12}}
\right)^{\f12}
\,.
\ee
On the other hand, the large spin asymptotics of spin network evaluations has been worked out using various methods, all based on coherent state techniques e.g. \cite{Dowdall:2009eg,Costantino},  and reads~:
\be
s_{\Gamma}
\underset{j_{\ell}\gg 1}\sim
\sum_{\Delta}
\sum_{\eps=\pm}
d_{\Delta,\eps}
e^{
\eps i\sum_{\ell\in\Delta}
j_{\ell}\theta_{\ell}
}
\,,
\ee
where the sum is over all immersions $\Delta$ in the Euclidean $\R^{3}$ space of the 2d triangulation dual to the graph $\Gamma$.
The pre-factors $d_{\Delta,\eps}$ decrease as a positive power of the spins. Their exact expressions (as  determinants of Hessian matrices) can be found in \cite{Dowdall:2009eg,Costantino}. It is not relevant here since they do not affect the leading order asymptotics of the sum over spins, and thus do not enter the computation of the asymptotic stationary points.
The angles $\theta_{\ell}$ are the dihedral angles between the two triangles linked by $\ell$ (or equivalently, sharing the edge dual to the link $\ell$), as illustrated on fig.~\ref{fig:neighbours}. These are 3d angles reflecting that the 2d triangulation is not flat (even if planar).
The exponent $\sum_{\ell\in\Delta} j_{\ell}\theta_{\ell}$ is actually the discrete equivalent of the integral over the surface of its extrinsic curvature, which enters the definition of the celebrated Regge action for discretized general relativity \cite{Regge:1961px}.

Putting these asymptotics together, we get a sum over immersions of the 2d triangulation and a sign $\eps$:
\be
s_{\Gamma}(\{j_\ell\}) 
\om_{\Gamma}(\{j_\ell\})
\underset{j_{\ell}\gg1}\sim
\sum_{\Delta,\eps}
f_{\Delta}^{(\eps)}(\{j_\ell\})\,e^{\phi_{\Delta}^{(\eps)}(\{j_\ell\})}
\,.
\ee
Setting aside the polynomial pre-factor $f_{\Delta}^{(\eps)}(\{j_{\ell}\})$, we focus on the exponential term:
\be
\phi_{\Delta}^{(\eps)}(\{j_{\ell}\})=
S_{R}(\{j_{\ell}\})+S_{2d}(\{j_{\ell}\})
\ee
with
\be
\left|\begin{array}{lcl}
S_{R}
&=&
\eps \sum_{\ell\in\Delta}
ij_{\ell}\theta_{\ell}
\,,\vspace*{2mm}\\
S_{2d}&=&
\f12\sum_{n}
\Big{[}
J_{n}\ln J_{n} -\sum_{\ell\ni n}(J_{n}-2j_{\ell})\ln(J_{n}-2j_{\ell})
\Big{]}\,.
\end{array}\right.
\nn
\ee
The notation $S_R$ is for the Regge action and $S_{2d}$ is for ``two-dimensional'' (and the motivation will be clear below). Since we have a sum over immersed triangulations $\Delta$, we can focus on one triangulation at a time. The existence of a stationary line for that triangulation will be enough to make the whole sum divergent. Poles will then be given by the set of stationary lines for all triangulations.

To solve the stationary point equation $\pp_{j_{\ell}}\phi=-2\ln Y_{\ell}$, we focus on a given link $\ell_{0}$, and its dual edge $e_{0}$, and gather all the terms depending on the spin $j_{0}$ carried by that link.
First, we have to differentiate the $S_{2d}$ term involving the spins around the two nodes sharing our considered link $\ell_{0}$. Around one of the node we denote the two additional spins $j_1$ and $j_2$, and around the other nodes we denote them $\tj_1, \tj_2$. Moreover we set $J = j_0 + j_1 + j_2$ and $\tJ = j_0 + \tj_1 + \tj_2$.
%
The variation of $S_{2d}$ with respect to the spin $j_{0}$ carried by the link $\ell_{0}$ is: 
\begin{widetext}
\be
\pp_{j_{0}}S_{2d}
=
\f12\ln\left[
\f{J(j_{1}+j_{2}-j_{0})}{(j_{0}+j_{1}-j_{2})(j_{0}+j_{2}-j_{1})}
\f{\tJ(\tj_{1}+\tj_{2}-j_{0})}{(j_{0}+\tj_{1}-\tj_{2})(j_{0}+\tj_{2}-\tj_{1})}\right]
\,.
\ee
\end{widetext}
Interpreting the spins as the dual edge lengths, a little trigonometry allows to express the above ratios of products of lengths in terms of the opposite angles of the two triangles sharing the edge $e_{0}$, as illustrated on fig.~\ref{fig:neighbours}
\be
\pp_{j_{0}}S_{2d}
=
-\ln\left(\tan\f\vphi2\tan\f{\tilde{\vphi}}2\right)
\,.
\ee
This term only sees the 2d intrinsic geometry of the triangulation, and is not sensitive to its embedding in the surrounding three-dimensional space.

Then, we have to differentiate the Regge action $S_{R}$ and, in particular, deal with the differential of the dihedral angles $\theta_{\ell}$ which enter it. They describe the extrinsic geometry of the 2d triangulation and reflect how it is embedded in the ambient 3d space.  Since we are dealing with a 2d triangulation, with the topology of a 2-sphere, embedded in the flat $\R^{3}$ Euclidean space, it is in fact a polyhedron, and satisfies the Schl\"afli identity \footnotemark{} \cite{luo}:
\be
\sum_{\ell\in\Delta}
j_{\ell}\,\rd\theta_{\ell}
=0
\,.
\ee
\footnotetext{
If one considers one point within the 2-sphere and draws lines from that inner point to all the vertices of the 2d triangulation, one obtains a 3d triangulation of the 3-ball, where every boundary triangle has been raised to a tetrahedron whose summit is that inner point. Then the Sch\"afli identity for the 2d triangulation simply results from the sum of the Schl\"afli identity for all the tetrahedra, since the whole construction lives in the flat $\R^{3}$ space.
}
This implies that the differential of $S_{R}$ with respect to $j_{0}$ is simply given by the dihedral angle between the planes of the two triangles sharing the edge $e_0$:
\be
\pp_{j_{0}}S_{R}
=
i\eps\theta_{0}
\,.
\ee
Putting the two contributions together, we conclude that a stationary spin configuration $\{j_{\ell}\}$ for the quantum gravity amplitude, thus satisfying $\pp_{j_{\ell}}\phi_{\Delta}^{(\eps)}+2Y_{\ell}=0$ for all links $\ell$, is given by the spins being the edge lengths of a 2d triangulations, dual to the graph $\Gamma$, immersed in the flat 3d space $\R^{3}$, such that the 2d and 3d angles of this triangulation satisfy  a compatibility equation with the $Y$-couplings:
\be
\label{zeroes}
\forall \ell\in\Delta
\,,\,\,
Y_{\ell}^{(0)}=
e^{i\eps\f{\theta_{\ell}}2}\,
\sqrt{\tan\f{\vphi_{\ell}}2\tan\f{{\tilde{\vphi}_{\ell}}}2}
\,,
\ee
where $\eps$ is a global sign.
%
Since this is a condition on the angles of the triangulation, it means that the stationary spin configuration $\{j_{\ell}\}$ is scale-invariant, i.e. one can arbitrarily rescale the spins by a global factor, $\{j_{\ell}\}\mapsto \{\lambda j_{\ell}\}$ with $\lambda\in\R_{+}$, so that we actually have a stationary line,  thus leading to a divergence of the sum over spins defining the quantum gravity amplitude $\cZ_{\Gamma}$. In turn, this means that the compatibility condition given for the $Y_{\ell}$'s actually gives zeroes of the loop polynomial, and thus a zeroes of the 2d Ising partition function:
\be
P_{\Gamma}[\{Y_{\ell}^{(0)}\}]
=0=
\cI_{\Gamma}(\{ \textrm{artanh}\,Y_{\ell}^{(0)}\})
\,.
\ee
This is the main result of the present letter: the formula above  in terms of the angles of dual 2d triangulations embedded in the flat Euclidean 3d space $\R^{3}$, gives zeroes of the inhomogeneous 2d Ising model partition function, on a planar trivalent graph.

At this point, it is not clear what proportion of the whole set of zeroes would be given by that formula. Indeed, on the one hand, assuming that both triangles angles $\vphi_{\ell},\tilde{\vphi}_{\ell}$ and dihedral angles $\theta_{\ell}$ are fixed (up to some finite degeneracy) by the edge lengths (obviously, up to a global scale), our geometric formula defines a $\R^{L-1}$ manifold of Ising zeroes, where $L$ is the number of links of the graph , and thus the number of edges of the dual triangulation. On the other hand, we have $L$ complex couplings $Y_{\ell}$ satisfying a single complex condition, that they are roots of the loop polynomials, thus leading to an expected $\C^{L-1}$ manifold of Ising zeroes. Following this logic, we expect that one would, at least, need to complexify our geometric formula in order to get a more thorough description of the Ising zeroes.

\section{Zeroes \& Critical Couplings}

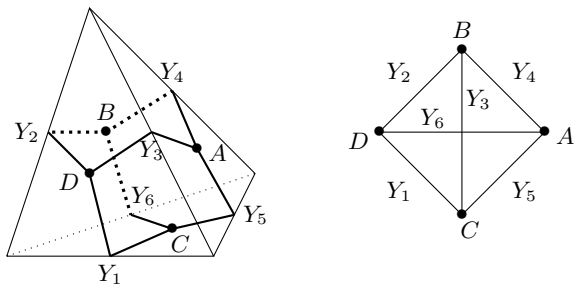
\begin{figure}[h!]

\centering

\begin{tikzpicture}[scale=1.1]

\coordinate(a) at (0,0) ;
\coordinate(ab) at (1.25,0) ;
\coordinate(b) at (2.5,0);
\coordinate(bc) at (1.75,1.5) ;
\coordinate(ac) at (.5,1.5);
\coordinate(c) at (1,3);
\coordinate(ad) at (1.5,.5);
\coordinate(bd) at (2.75,.5);
\coordinate(cd) at (2,2);
\coordinate(d) at (3,1);

\coordinate(C) at (2,.33);
\draw[thick] (C)--(ab);
\draw[thick] (C)--(bd);
\draw[thick] (C)--(ad);

\coordinate(A) at (2.3,1.3);
\draw[thick] (A)--(bc);
\draw[thick] (A)--(cd);
\draw[thick] (A)--(bd);

\coordinate(B) at (1.2,1.5);
\draw[very thick,dotted] (B)--(ac);
\draw[very thick,dotted] (B)--(ad);
\draw[very thick,dotted] (B)--(cd);

\coordinate(D) at (1,1);
\draw[thick] (D)--(ab);
\draw[thick] (D)--(bc);
\draw[thick] (D)--(ac);

\draw (a)--(b)--(c)--(a);
\draw[dotted] (a)--(d);
\draw (b)--(d)--(c);

\draw (A) node {$\bullet$} ++(.25,-0.05) node{$A$} ;
\draw (B) node {$\bullet$}++(0,.24) node{$B$};
\draw (C) node {$\bullet$}++(.1,-.18) node{$C$} ;
\draw (D) node {$\bullet$}++(-.25,-.1) node{$D$};

\draw(ab)node[below]{$Y_{1}$} ;
\draw(ac)node[left]{$Y_{2}$} ;
\draw(bc)node[below]{$Y_{3}$} ;
\draw(ad)++(0.15,.23) node{$Y_{6}$} ;
\draw(bd)node[right]{$Y_{5}$} ;
\draw(cd)node[above]{$Y_{4}$} ;


\coordinate(DD) at (4.5,1.5);
\coordinate(CC) at (5.5,.5);
\coordinate(BB) at (5.5,2.5);
\coordinate(AA) at (6.5,1.5);
\draw (AA) node {$\bullet$} ++(.25,-0.05) node{$A$} ;
\draw (BB) node {$\bullet$}++(0,.24) node{$B$};
\draw (CC) node {$\bullet$}++(.1,-.18) node{$C$} ;
\draw (DD) node {$\bullet$}++(-.25,-.1) node{$D$};
\draw (AA)--node[midway,above right]{$Y_{4}$}(BB);
\draw (AA)--node[midway,below right]{$Y_{5}$}(CC);
\draw (AA)--(DD);
\draw  (5.7,1.9) node {$Y_{3}$};
\draw (BB)--(CC);
\draw  (5.15,1.67) node {$Y_{6}$};
\draw (BB)--node[midway,above left]{$Y_{2}$}(DD);
\draw (CC)--node[midway,below left]{$Y_{1}$}(DD);

\end{tikzpicture}

\caption{
Tetrahedral graph $\Gamma$ (on the right) and dual tetrahedron $\Delta$ (on the left) defining a 2d triangulation of the 2-sphere. We label the graph nodes, or equivalently the triangles, $A,B,C,D$. We dressed the graph links, and equivalently their dual edges on the triangulations, with the couplings $Y_{\ell}=\tanh y_{\ell}$.
\label{fig:tetra}}

\end{figure}

The above formula for zeroes of the 2d Ising model has been proved explicitly in the case of the tetrahedral graph, dual to the simplest (non-degenerate) 2d triangulation of the 2-sphere. Indeed, it was shown in \cite{Bonzom:2019dpg} that the parametrization \eqref{zeroes} above in terms of triangle and dihedral angles are zeroes of the loop polynomial on the tetrahedron, explicitly:
\beq
P_{\Gamma}
[Y_{1},..,Y_{6}]
&=&
1
+Y_{1}Y_{3}Y_{4}Y_{6}+Y_{3}Y_{2}Y_{6}Y_{5}+Y_{2}Y_{1}Y_{5}Y_{4}
\nn\\
&&
+Y_{1}Y_{2}Y_{6}+Y_{1}Y_{5}Y_{3}+Y_{4}Y_{2}Y_{3}+Y_{4}Y_{5}Y_{6}
\,,\nn
\eeq
where the edges are labelled according to fig.\ref{fig:tetra}.
This polynomial depends on 6 variables $Y_{i}$, so one should expect a ten-dimensional space of roots.  However, our formula depends only on 5 parameters (6 edge lengths up a global scale, or equivalently, 6 dihedral angles with a vanishing Gram matrix determinant). It was thus conjectured in \cite{Bonzom:2019dpg} that the whole space of zeroes of that loop polynomial is given by a suitable complex continuation of the formula \eqref{zeroes}. Here we extend this conjecture to arbitrary planar trivalent graphs and their dual polyhedra.

\medskip

Although it would be interesting to work out the case of more complicated beyond the tetrahedral graph, and in particular, find a direct proof (or disproof) of our geometric formula for Ising zeroes, it is also relevant to consider the thermodynamical limit, i.e. when we take the limit of an infinite graph. Then the zeroes of the partition function a priori make correlations diverge and, thus, should be compared to the critical couplings.

\begin{figure}[ht!]
\begin{center}
\includegraphics[height=30mm]{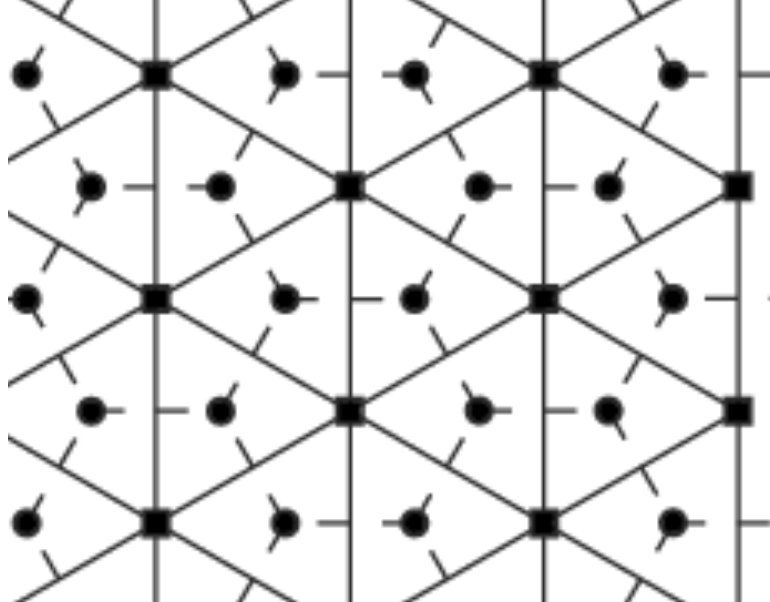}
\caption{
Honeycomb lattice and its dual triangulation
\label{fig:honeycomb}}
\end{center}
\end{figure}

Let us consider the homogeneous Ising model on the honeycomb lattice, which is indeed a trivalent graph. It is dual to a 2d triangulation made of identical equilateral triangles, as shown on fig.\ref{fig:honeycomb}. Our Ising zeroes formula reproduces exactly the known critical coupling (which can be found by exact techniques, such as the star-triangle relation \cite{Baxter:1982zz}):
\be
Y^{(0)}=\tan\f\pi6=\f1{\sqrt{3}}=Y_{c}\,.
\ee

\begin{figure}[ht!]
\begin{center}
\includegraphics[height=50mm]{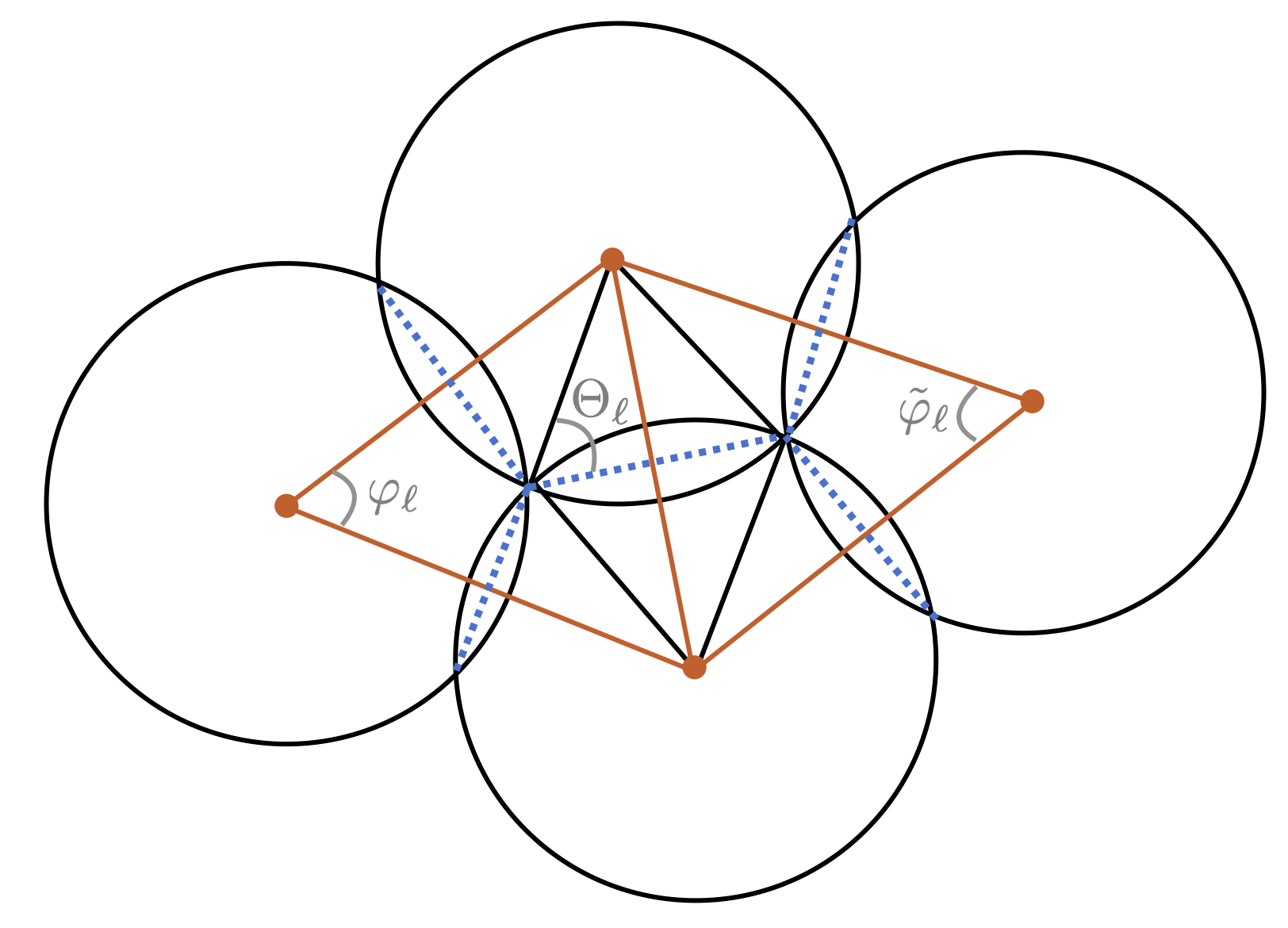}
\caption{
A isoradial graph (in {\color{blue}dotted blue}) is built from intersecting circles, all with the same radius. The nodes of the graph are located at the intersection of the circles, so the faces of the graph are inscribed in the circles. The vertices of the dual triangulation (in {\color{red}red}) are the center of the circles. Focussing on a graph link $\ell$, the lines between its source and target nodes and the end vertices of its dual edge form a rhombus, whose half-angle is noted $\Theta_{\ell}$. basic geometry shows that it is equal to both the opposite 2d angles $\vphi_{\ell}$ and $\tilde{\vphi}_{\ell}$.
\label{fig:isoradial}}
\end{center}
\end{figure}

An inhomogenous generalisation of the honeycomb lattice are isoradial graphs (see e.g. \cite{David:2013nta}), that is graphs drawn in the $\R^{2}$ plane such that all its faces are inscribable in circles with the same radius. As illustrated on fig.\ref{fig:isoradial}, the vertices of the dual triangulations are the center of those circles. In that case, dihedral angles  obviously vanish, since the triangulation is plain flat, and the two opposite angles for each edge are equal to its half-rhombus angle $\Theta_{\ell}$.
Our formula thus gives
\be
Y^{(0)}_{\ell}=\tan\f{\Theta_{\ell}}2\,,
\ee
which is exactly the critical coupling value $Y_{c}=\tanh y_{c}$ derived by discrete holomorphicity or by dimer models \cite{Boutillier_2018}. Our quantum gravity formula thus proposes a substantial generalization of Ising criticality beyond flat graphs and isoradiality.

\section{Higher Valence and Circle Patterns}

We would like to conclude our analysis by showing that, assuming that our formula is indeed correct for 3-valent planar graphs, then it straightforwardly extends to planar graphs with arbitrary valence, in terms of the geometry of circle patterns embedded in the flat 3d space.

\begin{figure}[ht!]
\begin{center}
\begin{tikzpicture}[scale=3.5]

\coordinate (A1) at (0,0);
\coordinate (A2) at (0.5,-0.1);
\coordinate (A3) at (.65,.4);
\coordinate (A4a) at (.26,.36);
\coordinate (A4b) at (.39,.42);


\draw (A1)--  node[pos=0.5,inner sep=0pt](m12){} (A2);
\draw (A2)--  node[pos=0.5,inner sep=0pt](m23){}(A3) --node[pos=0.5,inner sep=0pt](m34){} (A4b) --  node[pos=0.5,inner sep=0pt](m4){} (A4a) -- node[pos=0.5,inner sep=0pt](m14){} (A1);
\fill[fill=black,fill opacity=0.05]  (A1) -- (A2) -- (A3) -- (A4b)-- (A4a);

\coordinate (A5) at (.95,0);
\coordinate (A6) at (1.1,.65);
\draw (A2) --  node[pos=0.5,inner sep=0pt](m25){} (A5) --  node[pos=0.5,inner sep=0pt](m56){} (A6) --  node[pos=0.5,inner sep=0pt](m36){} (A3);
\fill[fill=black,fill opacity=0.05]  (A2) -- (A5) -- (A6) -- (A3);

%

\coordinate (A7) at (.57,.8);
\draw (A3) --  node[pos=0.5,inner sep=0pt](m37){} (A7) -- node[pos=0.5,inner sep=0pt](m67){}(A6);
\fill[fill=black,fill opacity=0.05]  (A3) -- (A7) -- (A6);
\draw (A4b) -- node[pos=0.5,inner sep=0pt](m47){}(A7);
\fill[fill=black,fill opacity=0.05]  (A4a) -- (A7) -- (A3);

\coordinate (A8) at (.93,1.05);
\coordinate (A9) at (.62,1.2);
\draw (A6) --node[pos=0.5,inner sep=0pt](m68){} (A8) -- node[pos=0.5,inner sep=0pt](m89){}(A9)  -- node[pos=0.5,inner sep=0pt](m79){}(A7);
\fill[fill=black,fill opacity=0.05]  (A6) -- (A8) -- (A9) -- (A7);

%

\coordinate (A10) at (-0.1,.75);
\draw (A9) -- node[pos=0.5,inner sep=0pt](m910){}(A10) -- node[pos=0.5,inner sep=0pt](m710){}(A7);
\fill[fill=black,fill opacity=0.05] (A9) -- (A10) -- (A7);
\draw (A10) -- node[pos=0.5,inner sep=0pt](m410){}(A4a);
\fill[fill=black,fill opacity=0.05] (A4a) -- (A10) -- (A7);
\draw (A10) -- node[pos=0.5,inner sep=0pt](m110){}(A1);
\fill[fill=black,fill opacity=0.05] (A4a) -- (A10) -- (A1);

\coordinate (a1) at (.73,.65);
\draw[dotted] (A8) -- node[pos=0.5,inner sep=0pt](N18){}(a1) -- node[pos=0.5,inner sep=0pt](N15){}(A5);
\coordinate (a2) at (.33,.6);
\draw[dotted] (A9) -- node[pos=0.5,inner sep=0pt](N19){}(a1) -- node[pos=0.5,inner sep=0pt](n12){}(a2) -- node[pos=0.5,inner sep=0pt](N210){}(A10);
\coordinate (a3) at (.55,.3);
\draw[dotted] (a1) --node[pos=0.5,inner sep=0pt](n13){} (a3) -- node[pos=0.5,inner sep=0pt](n23){}(a2);
\draw[dotted] (A1) -- node[pos=0.5,inner sep=0pt](N13){}(a3) --node[pos=0.5,inner sep=0pt](N35){} (A5) ;

\coordinate (B1) at (.4,.15);
\draw (B1) node{\blue\tiny{$\bullet$}};
\coordinate (B3) at (.1,.35);
\draw (B3) node{\blue\tiny{$\bullet$}};
\coordinate (B4) at (.33,.92);
\draw (B4) node{\blue\tiny{$\bullet$}};
\coordinate (b1) at (.75,1);
\draw[opacity=.2] (b1) node{\blue\tiny{$\bullet$}};
\coordinate (b123) at ($.33*(a1)+.33*(a2)+.33*(a3)$);
\draw[opacity=.2] (b123) node{\blue\tiny{$\bullet$}};
\coordinate (b1235) at ($.25*(a3)+.25*(A1)+.25*(A5)+.25*(A2)$);
\draw[opacity=.2] (b1235) node{\blue\tiny{$\bullet$}};
\coordinate (b12) at ($.25*(a1)+.25*(a2)+.25*(A9)+.25*(A10)$);
\draw[opacity=.2] (b12) node{\blue\tiny{$\bullet$}};
\coordinate (b158) at ($.25*(a1)+.25*(A5)+.25*(A8)+.25*(A6)$);
\draw[opacity=.2] (b158) node{\blue\tiny{$\bullet$}};
\coordinate (b135) at ($.33*(a1)+.33*(A5)+.33*(a3)$);
\draw[opacity=.2] (b135) node{\blue\tiny{$\bullet$}};
\coordinate (b23) at ($.25*(a3)+.25*(A1)+.25*(a2)+.25*(A10)$);
\draw[opacity=.2] (b23) node{\blue\tiny{$\bullet$}};

\coordinate (B5) at ($.33*(A4b)+.33*(A7)+.33*(A10)$);
\draw (B5) node{\blue\tiny{$\bullet$}};
\coordinate (B6) at ($.33*(A4b)+.33*(A7)+.33*(A3)$);
\draw (B6) node{\blue\tiny{$\bullet$}};
\coordinate (B7) at ($.33*(A6)+.33*(A7)+.33*(A3)$);
\draw (B7) node{\blue\tiny{$\bullet$}};
\coordinate (B8) at ($.25*(A2)+.25*(A5)+.25*(A3)+.25*(A6)$);
\draw (B8) node{\blue\tiny{$\bullet$}};
\coordinate (B9) at ($.25*(A6)+.25*(A7)+.25*(A8)+.25*(A9)$);
\draw (B9) node{\blue\tiny{$\bullet$}};


\draw[blue] (B5)--(m4)--(B1);
\draw[blue] (B1)--(m12);
\draw[blue] (B1)--(m23);
\draw[blue] (B1)--(m14);
\draw[blue] (B1)--(m34);


\draw[blue] (B3)--(m14);
\draw[blue] (B3)--(m110);
\draw[blue] (B3)--(m410);

\draw[blue] (B4)--(m79);
\draw[blue] (B4)--(m710);
\draw[blue] (B4)--(m910);

\draw[blue] (B5)--(m47);
\draw[blue] (B5)--(m410);
\draw[blue] (B5)--(m710);

\draw[blue] (B6)--(m47);
\draw[blue] (B6)--(m34);
\draw[blue] (B6)--(m37);

\draw[blue] (B7)--(m67);
\draw[blue] (B7)--(m36);
\draw[blue] (B7)--(m37);

\draw[blue] (B8)--(m23);
\draw[blue] (B8)--(m25);
\draw[blue] (B8)--(m56);
\draw[blue] (B8)--(m36);

\draw[blue] (B9)--(m67);
\draw[blue] (B9)--(m68);
\draw[blue] (B9)--(m79);
\draw[blue] (B9)--(m89);

%

\draw[blue, opacity=.2] (b123)--(n12);
\draw[blue, opacity=.2] (b123)--(n23);
\draw[blue, opacity=.2] (b123)--(n13);

\draw[blue, opacity=.2] (b1235)--(m12);
\draw[blue, opacity=.2] (b1235)--(m25);
\draw[blue, opacity=.2] (b1235)--(N35);
\draw[blue, opacity=.2] (b1235)--(N13);

\draw[blue, opacity=.2] (b135)--(N35);
\draw[blue, opacity=.2] (b135)--(n13);
\draw[blue, opacity=.2] (b135)--(N15);

\draw[blue, opacity=.2] (b158)--(N15);
\draw[blue, opacity=.2] (b158)--(N18);
\draw[blue, opacity=.2] (b158)--(m56);
\draw[blue, opacity=.2] (b158)--(m68);

\draw[blue, opacity=.2] (b1)--(m89);
\draw[blue, opacity=.2] (b1)--(N19);
\draw[blue, opacity=.2] (b1)--(N18);

\draw[blue, opacity=.2] (b12) --(n12);
\draw[blue, opacity=.2] (b12) --(N210);
\draw[blue, opacity=.2] (b12) --(N19);
\draw[blue, opacity=.2] (b12) --(m910);

\draw[blue, opacity=.2] (b23) --(m110);
\draw[blue, opacity=.2] (b23) --(n23);
\draw[blue, opacity=.2] (b23) --(N13);
\draw[blue, opacity=.2] (b23) --(N210);

\end{tikzpicture}
\caption{2d cellular complex, with the topology of a 2-sphere, made of polygons glued together, and its dual graph in {\color{blue} blue}.
\label{fig:2sphere}}
\end{center}
\end{figure}
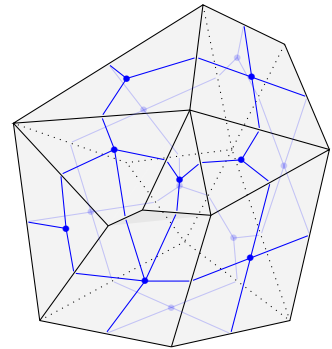

Indeed, while a 3-valent node is naturally dual to a triangle, a node with higher valence is dual to a polygon with as many edges as links connected to the node. Then a planar graph with nodes of arbitrary valence (higher or equal to 3) can be embedded in 3d space as dual to a 2d cellular complex with the topology of a 2-sphere, as illustrated on fig.\ref{fig:2sphere}.

\begin{figure}[ht!]
\begin{center}
\includegraphics[width=85mm]{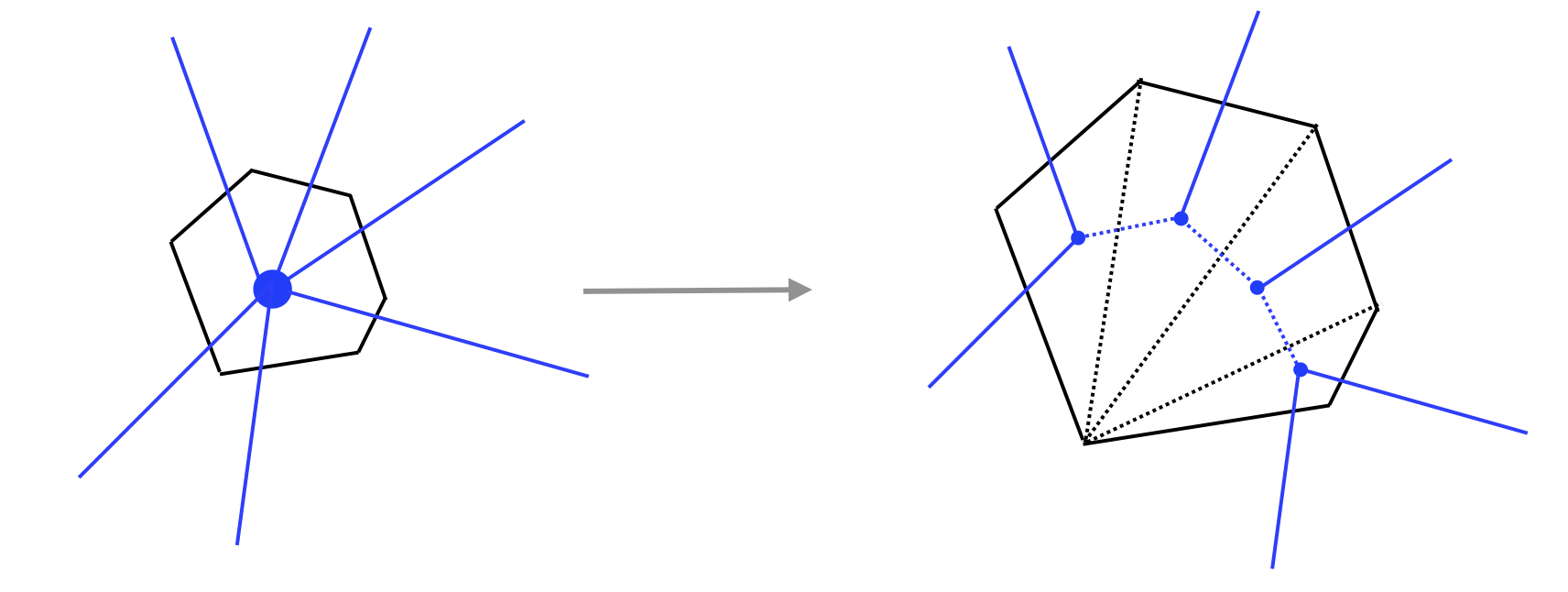}
\caption{
Decomposition of a polygon into triangles and its dual unfolding of a higher valent node into a tree of 3-valent nodes. Internal triangulation edges ad their dual internal links are in dotted black and {\color{blue} dotted blue}.
\label{fig:unfold}}
\end{center}
\end{figure}

A polygon can be decomposed into triangles by choosing one of its vertices and drawing lines from it to all its other vertices. As drawn on fig.\ref{fig:unfold}, this unfolds the corresponding graph node into a tree of 3-valent nodes.
We will refer to the new links between these 3-valent nodes as {\it internal links}.
Doing so at every node turns the original graph $\Gamma$ into a planar 3-valent graph $\bGamma$, to which we can now apply our Ising zeros' formula.
The key point is that the loop polynomial on the original graph is (obviously) exactly equal to the the loop polynomial on the unfolded 3-valent graph, if setting all the Ising coupling $Y_{\ell}$ on the internal links to 1:
\be
P_{\Gamma}[\{Y_{\ell}\}_{\ell\in\Gamma}]
=
P_{\bGamma}[\{Y_{\ell}\}_{\ell\in\Gamma},\{Y_{\ell}=1\}_{\ell\textrm{ internal}}]
\,.
\ee

\begin{figure}[ht!]
\begin{center}
\includegraphics[height=50mm]{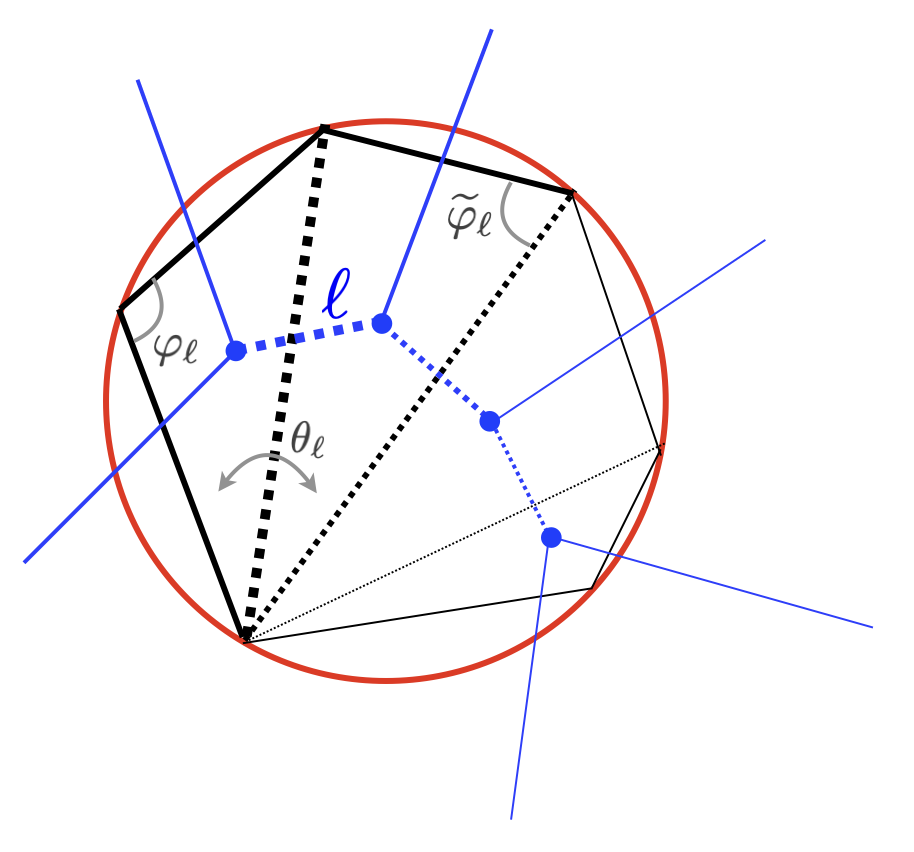}
\caption{
Two neighbouring triangles in a polygon inscribed in circle. The geometric coupling $Y_{\ell}$ on the {\it internal} link $\ell$ between the two triangles is automatically equal to 1, as the dihedral angle vanishes $\theta_{\ell}=0$ and the opposite triangle angles sum to $\vphi_{\ell}+\widetilde{\vphi}_{\ell}=\pi$.
\label{fig:circle}}
\end{center}
\end{figure}

Therefore we get zeros of the loop polynomial on the graph $\Gamma$ from the loop polynomial on the extended graph $\bGamma$. The latter has only 3-valent nodes and we can apply our geometric ansatz \eqref{zeroes} for Ising zeros.
We thus need to understand the geometrical meaning of setting $Y_{\ell}=1$ on a link. Looking at the formula \eqref{zeroes}, one sees that the coupling $Y_{\ell}$ is equal to 1 if and only if its phase given by the dihedral angle vanishes, $\theta_{\ell}=0$, and the product of the tangent of the half triangle angle is equal to 1, i.e. if the sum of the opposite triangle angles is equal to $\vphi_{\ell}+\widetilde{\vphi}_{\ell}=\pi$. This is automatically the case for two neighbouring triangles if the polygon is inscribed in a circle, as drawn on fig.\ref{fig:circle}. This provides a  natural geometric (sufficient) condition for gluing 3-valent Ising vertices into a higher valent vertex.

\begin{figure}[ht!]
\begin{center}
\includegraphics[width=90mm]{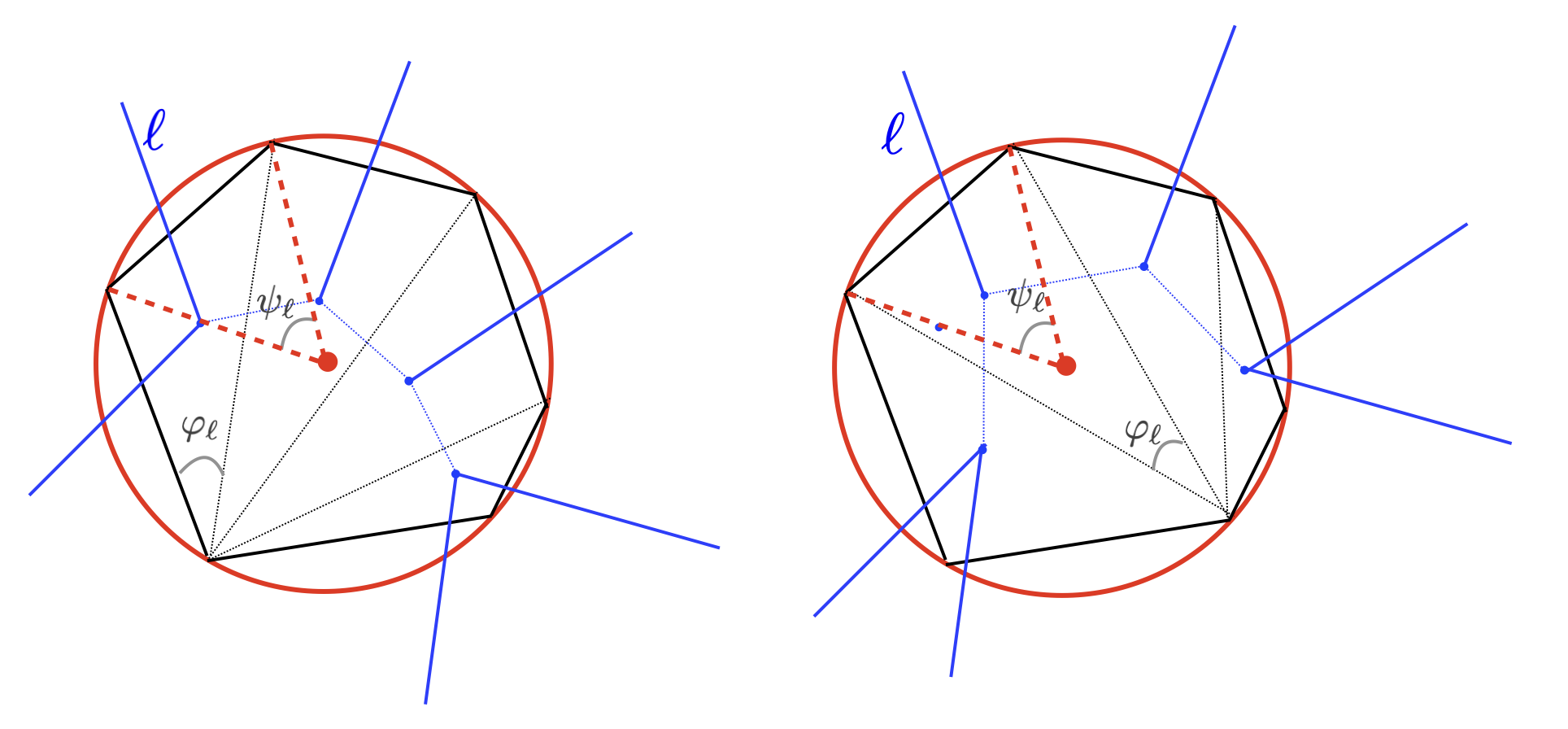}
\caption{
Triangle angles and center angles for the Ising zeros' geometric ansatz \eqref{highervalence} on embedded circle patterns. Whatever the decomposition of the polygon into triangles, the center angle $\psi_{\ell}$ for a graph link $\ell$ (not an internal link) is always twice the opposite triangle angle $\varphi_{\ell}$. Thus the generalized formula \eqref{highervalence} for higher valent graphs naturally follows from the geometric ansatz \eqref{zeroes} for 3-valent graphs and their dual 2d triangulations.
\label{fig:circleangle}}
\end{center}
\end{figure}

So let us assume that every polygon is inscribed in a circle, as drawn on fig.\ref{fig:circle}. This is called a {\it circle pattern}. Work on Ising critical couplings on circle patterns was done e.g. in \cite{lis2019circle} (see also \cite{chelkak2022ising}).
The extra ingredient of the present analysis is that the circle pattern is not drawn in the flat 2d space but embedded in the flat 3d space, and that the dihedral angles between the polygon planes play the key role of determining the phases of the Ising couplings.
Indeed, our formula for the Ising zeros of the 3-valent graphs leads directly to a generalized formula on the graph $\Gamma$ with arbitrary valence. The ansatz for the couplings on the links of the graph (and not on the internal link, which have been set to 1) is:
\be
\label{highervalence}
\forall \ell\in\Delta
\,,\,\,
Y_{\ell}^{(0)}=
e^{i\eps\f{\theta_{\ell}}2}\,
\sqrt{\tan\f{\psi_{\ell}}4\tan\f{{\widetilde{\psi}_{\ell}}}4}
\,,
\ee
where $\psi_{\ell}$ and $\tilde{\psi}_{\ell}$ are now the angles at the center of the two circles sharing the edge dual to the link $\ell$, as drawn on fig. \ref{fig:circleangle}. Due to basic trigonometry, these are automatically equal to the opposite triangle angle whatever the chosen decomposition of the polygon into triangles.
The angle $\theta_{\ell}$ remains the dihedral angle between the normal vectors to the two polygons linked by $\ell$ (or equivalently sharing its dual edge).

This shows that the geometric formula for Ising zeros from quantum gravity naturally extends from 3-valent graphs (dual to triangulations) to graphs with nodes of arbitrary higher valence embedded in circle patterns, but also underlines the potentially interesting role of circle patterns on the boundary of the Ponzano-Regge path integral for 3d quantum gravity.

\section*{Outlook}

To summarize, we have used the BCL duality between 3d quantum gravity and the 2d inhomogeneous Ising model, put in light in \cite{Bonzom:2015ova} and based on Westbury formula  \cite{Westbury1998}, to design a new method to find the zeros of the Ising model. This duality relates the Ponzano-Regge path integral amplitudes for 3d quantum gravity for a bounded region of space-time to the 2d Ising model on the boundary surface, where the Ising couplings are given in terms of the coherent state parameters determining the quantum state of geometry on the boundary. More precisely, the Ponzano-Regge state-sum is equal, up to factors, to the inverse square of the Ising partition function. This allowed us to describe the 2d Ising zeroes in terms of the semi-classical regime of quantum gravity, using asymptotic saddle points of the Ponzano-Regge amplitudes at large spins. This led us to a new formula \eqref{zeroes} for the Ising zeroes. Remarkably, it is consistent with the formula for Ising's critical couplings on isoradial graphs and it extends the latter into an expression for the critical couplings in terms of the geometry of 2d triangulations embedded in the flat 3d space.
Moreover, we were able to extend our geometric ansatz from 3-valent graphs, dual to 2d triangulations, to graphs with arbitrary valence, dual to general 2d circle patterns.

This opens a new door between quantum gravity and statistical physics, with research in quantum gravity bringing a new perspective and original methods to tackle statistical physics from a geometrical standpoint.
%
%
To go further in this direction, it would first be enlightening to make the proof of this Ising zeroes formula mathematically rigorous and, in particular, to find a direct proof, as done for instance in \cite{Bonzom:2019dpg} for the special case of the tetrahedral graph.
One should also investigate if the whole set of Ising zeroes, i.e. the roots of the loop polynomial, can be parametrized by suitably complexifying our formula for ``real'' zeroes.
On the statistical physics side, this might allow in particular to have a representation of the low temperature -- high temperature duality of the 2d Ising model. Indeed, it was already found in \cite{Bonzom:2019dpg} that the ``real'' formula for the tetrahedral graph was not stable under the low T -- high T duality transformation and identifying a consistent complexification would allow to address this issue. 
On the quantum gravity front, such complexification of the geometry echoes recent works on the analysis of the crucial contribution of complex saddle points to spinfoam path integral amplitudes \cite{Han:2021kll}.
It will be interesting to understand how these aspects are related to each other.

Then, widening the scope of investigation, one could seek similar duality formulas, for example from non-linear supersymmetric extensions as proposed in \cite{Bonzom:2015ova}, or simply by pushing the Ising model to Potts models, or by investigating what happens by curving the 2d triangulations or the ambient 3d space.
At the end of the day, we hope that this bridge can be developed into a fruitful interface between quantum gravity and statistical physics, searching for geometrical formulas for statistical physics and condensed matter models from quantum gravity's holographic dualities at finite distance.

\section*{Acknowledgement}

The authors would like to thank Marcin Lis for pointing out works
studying the Ising model on circle patterns.

%
%




\bibliographystyle{bib-style}
\bibliography{TQFT}

\end{document}